# Zel´dovich Λ and Weinberg's Relation: An Explanation for the Cosmological Coincidences

ANTONIO ALFONSO-FAUS

*E.U.I.T. Aeronáutica Plaza cardenal Cisneros s/n, 8040 Madrid, SPAIN*
*e-mail: aalfonsofaus@yahoo.es*

**Abstract.** In 1937 Dirac proposed the large number hypothesis (LNH). The idea was to explain that these numbers were large because the Universe is old. A time variation of certain "constants" was assumed. So far, no experimental evidence has significantly supported this time variation. Here we present a simplified cosmological model. We propose a new cosmological system of units, including a cosmological Planck's constant that "absorbs" the well known large number $10^{120}$. With this new Planck's constant no large numbers appear at the cosmological level. They appear at lower levels, e.g. at the quantum world. We note here that Zel'dovich formula, for the cosmological constant Λ, is equivalent to the Weinberg's relation. The immediate conclusion is that the speed of light c must be proportional to the Hubble parameter H, and therefore decrease with time. We find that the gravitational radius of the Universe and its size are one and the same constant (Mach's principle). The usual cosmological Ω´s parameters for mass, lambda and curvature turn out to be all constants of order one. The anthropic principle is not necessary in this theory. It is shown that a factor of $10^{61}$ converts in this theory a Planck fluctuation (a quantum black hole) into a cosmological quantum black hole: the Universe today. General relativity and quantum mechanics give the same local solution of an expanding Universe with the law $a(t) \approx const.\ t$. This constant is just the speed of light today. Then the Hubble parameter is exactly $H = a(t)'/a(t) = 1/t$.

## 1. Introduction.

Much theoretical and experimental work has been done in the past using the hypothesis that the gravitational "constant" G varies with time. In 1937 Dirac proposed his large number hypothesis (LNH) suggested by the numerical coincidences of two large numbers: the ratio of electric and gravitational forces between an electron and a proton, which is of the order of $10^{40}$, and the ratio of the age of the Universe to the time light takes to travel the size of a fundamental particle. Dirac generalized this result to say that any number which is a power of $10^{40}$ will be time dependent to the same power, which constitutes his LNH. He kept all quantum properties as constant and left G to vary as 1/t. For the number of particles in the Universe, $N_p \approx 10^{79}$, this hypothesis implies $N_p \approx t^2$. The results of many experimental observations do not support this hypothesis, Uzan (2003).

Besides, theoretically we know that the number of photons in the Universe $N_{ph}$ is of the order of the third power of the ratio R/λ, R ≈ $10^{28}$ cm. the size of the Universe, and λ the typical wavelength of a photon in the cosmic microwave background, a blackbody radiation with typical λ ≈ 0.1 cm. Quantum mechanics imposes that R and λ be proportional so that $N_{ph}$ ≈ $10^{87}$ = constant. The ratio $N_{ph}/N_p$ is not observed to vary with time (lithium content in the Universe). And no creation of particles is observed either. The conclusion is that $N_p$ appears to be constant and therefore the Dirac LNH is not confirmed by experiments. He expressed the idea that very large dimensionless universal constants cannot be pure mathematical numbers, and therefore that they must not occur in the basic laws of physics. Then he considered that this numbers were large because they are time varying, and the Universe is old. But this idea may be pursued in a different way, introducing a cosmological Planck's constant. Then we will see that at a cosmological scale all dimensionless universal constants are of order one, which basically is Dirac's first idea. Therefore at a universal scale there are no large numbers at all, as Dirac thought. This means that the dimensionless cosmological numbers $\Omega_m$, $\Omega_\Lambda$ and $\Omega_k$, as used in the Einstein cosmological equations, are all constants of order one. So is the cosmological constant in the cosmological units we choose. We conclude that the large numbers come in when considering smaller scales, smaller than the whole Universe. In particular they come in at the quantum scale.

In 1967 Zel'dovich published an expression for the energy density of vacuum. He obtained a formula for the Λ term, the cosmological constant, in terms of G, a fundamental particle mass m, and Planck's ℏ. Later on, in 1972, Weinberg found a relation between a fundamental particle mass m, G and ℏ, the same as in the Zel'dovich expression, but including now the speed of light c and the Hubble parameter H. The inclusion of c and H in Weinberg's relation imposes that c, the speed of light, must be proportional to the Hubble parameter H, as presented elsewhere by Alfonso-Faus (2008). This forces the speed of light to decrease linearly with cosmological time, the same as H.

We accept Mach's principle in the form saying that the gravitational radius of the Universe is of the same order of magnitude as its size (approximately given by the product ct, t the age of the Universe). We find that the Einstein's cosmological equations have a solution in complete agreement with the current values of the cosmological parameters. The introduction of a cosmological Planck's constant avoids the problem of the cosmological constant and represents one more step in the long way to integrate general relativity with quantum mechanics. Then, we find that Zel'dovich expression and Weinberg's relation, which is one and the same thing, are both satisfied at the cosmological scale as well as at the quantum scale. Since we have proven, Alfonso-Faus (2008), that this last relation is a result of the combined application of the conservation of momentum and the universal gravitation (Newton's laws), so is the Zel'dovich expression.

**2. System of units for a cosmological scale. The action integral.**

Einstein's field equations can be derived from an action integral following the Least Action Principle. In standard general relativity one has for the action integral, Weinberg (1972):

$$A = I_G + I_M \tag{1}$$

$$A = -\frac{c^3}{16\pi G}\int RR(g)^{1/2}d^4x + I_M$$

where $I_M$ is the matter action and $I_G$ the gravitational term. Then one obtains the field equations

$$G^{\mu\upsilon} = 8\pi\left(\frac{G}{c^4}\right)T^{\mu\upsilon} + \Lambda g^{\mu\upsilon} \tag{2}$$

We assume a space-time metric and use the Robertson-Walker model that satisfies the Weyl postulate and the cosmological principle. Einstein's equations (2) follow from the Action (1) provided that the variation of the coefficient in the integral in equation (1) is zero. This condition is,

$$\frac{c^3}{16\pi G} = const \tag{3}$$

We see that the assumption of a time varying G must include a time varying c to preserve the form of the field equations, and vice versa.

The equation (3) strongly suggests a specific link between mass and time. This is

$$\frac{c^3}{G} \approx 4 \times 10^{38}\, grams/\sec = const \tag{4}$$

which is of the order of the ratio of the mass of the observable Universe to its age. It is also of the order of the ratio of Planck's mass to Planck's time Mass and time seem to be intimately related. On the other hand, the action for a free material point is:

$$A = -mc \int ds \qquad (5)$$

To preserve standard mechanics we make the coefficient in (5), the momentum mc, constant independent from the cosmological time. Then

$$mc = const \qquad (6)$$

To preserve the validity of special relativity we assume v/c = constant, independent of cosmological time, so that from (6) this is equivalent to say that mv = constant, i.e., that linear momentum is conserved.

With the constancies expressed in (4) and (6), general relativity is preserved in the sense that the Einstein's field equations can be derived from the action integral, and of course the Newtonian mechanics too. If one keeps the expressions (4) and (6) as constants then time variations of some of the fundamental constants, G, c and masses, are allowed because the laws of physics as we know today are preserved.

Within a time interval of only a few years, Zel'dovich's (1967) expression and Weinberg's (1972) relation were published. They are in fact the same if the speed of light c is proportional to the Hubble parameter H. Since H is a cosmological parameter so is c, and both are time varying inversely proportional to the age of the Universe. This is a cosmological point of view. The speed of light decreases linearly with time as 1/t . It is evident that, with such a law for the speed of light, the size of the Universe (of the order of ct) is constant and therefore there is no "absolute" expansion.

The expression of Zel'dovich (1967), is as follows

$$\Lambda = 8\pi \frac{G^2 m^6}{\hbar^4} \qquad (7)$$

We are taking Planck's constant as a true universal constant. We are lead to this conclusion following the evidence of radioactive materials: the lifetime for the beta decay, Eisberg (1961), is proportional to the 7$^{th}$ power of ħ. The observed constancy of this time strongly imposes the constancy of ħ. Of course the constancy of angular momentum also imposes the constancy of ħ. On the other hand from (4) and (6) the product $Gm^3$ is a theoretical universal constant. Otherwise we could not derive Einstein's field equations from the action principle. Then from (7) the cosmological constant Λ is also a true universal constant. Following the order of magnitude of all the members of Einstein's cosmological equations, its value is of the order of $\Lambda/8\pi \approx 1/(ct)^2$. Again ct must be a constant since so is Λ. And if we choose as the *unit of length* the value ct ≈ $10^{28}$ cm for the cosmological scale, then the cosmological constant Λ is of order one. We can take as the *unit of linear momentum*, at the cosmological scale, the value of Mc, where M is the mass of the Universe. We then have two cosmological units: ct = 1 and Mc = 1. Its combination gives M = t (the mass-boom, Alfonso-Faus, 2008). We also have from (4) the choice of making $c^3$ = G. Summarizing we have the choice of cosmological units as follows:

$$ct=1; Mc=1; (combining\, M=t); c^3=G \quad (combining\ \frac{GM}{c^2}=1) \qquad (8)$$

Hence, the gravitational radius of the Universe is a constant (unit of length in this system) and equal to ct, the form of Mach's principle. The unit of angular momentum, at the cosmological scale, is

$$Mc(ct) = \hbar_c = 1 \qquad (9)$$

where $\hbar_c$ is the cosmological Planck's constant, equal to one. The cosmological counterpart of Zel'dovich relation (7) is now

$$\Lambda = 8\pi \frac{G^2 M^6}{\hbar_c^4} = 8\pi \frac{(Mc)^6}{\hbar_c^4} \qquad (10)$$

and using Mc = 1 and $\hbar_c$ = 1 we get $\Lambda/8\pi$ = 1. Then, it is confirmed that the cosmological constant being of order $1/(ct)^2$ with ct = 1 is also of order one, a universal constant. Zel'dovich relation (7) is not only valid at the quantum scale (using m and ħ) but also at the cosmological scale (using M and $\hbar_c$).

Expressing the gravitational radius of the Universe and its cosmological Compton wavelength, in a similar way as Planck's units in terms of G, c and ℏ_c, we get

$$\frac{GM}{c^2} = \frac{\hbar_c}{Mc} = ct = 1 \qquad (11)$$

We can interpret this by saying that the Universe is a cosmological quantum black hole of mass M, length ct, and time its age t. The large number $10^{61}$ converts Planck's fluctuation, mass length and time, into the cosmological one, i.e., the Universe today.

Weinberg's (1972) relation, predicting the Pion mass m, is given by

$$m^3 = A\frac{\hbar^2 H}{Gc} \qquad (12)$$

where A is a numerical constant close to one. Comparing this with (7) it is evident that c must be proportional to H. This conclusion has far reaching consequences, as seen in the next section. It has been derived elsewhere, Alfonso-Faus (2008), as follows:

$$c = HL \qquad (13)$$

where L is the constant of the order of the present size of the Universe L = ct ≈ $10^{28}$ cm. Of course Weinberg's relation (12) also holds for M, instead of m, and ℏ_c, instead of ℏ. We have generalized Zel'dovich and Weinberg's formulation to the cosmological level.

## 3. The Constancy of the Ω Cosmological Parameters

The dimensionless cosmological parameters $\Omega_m$, $\Omega_\Lambda$ and $\Omega_k$ are defined as

$$\Omega_m = \frac{8\pi}{3}\frac{G\rho}{H^2} = \frac{8\pi}{3}\frac{GM}{2\pi^2 L^3}\cdot\frac{1}{H^2} = \frac{4}{3\pi}\frac{GM}{c^2}\cdot\frac{1}{L} \approx 1$$

$$\Omega_\Lambda = \frac{\Lambda c^2}{3H^2} = \frac{\Lambda L^2}{3} \approx 1 \qquad (14)$$

$$\Omega_k = \frac{kc^2}{a^2}\cdot\frac{1}{H^2} \approx k = 1$$

The first relation in (14), as pointed out by Weinberg (1972), gives

$$G\rho t^2 \approx 1 \tag{15}$$

This is the same as

$$\frac{GM}{c^2} \approx ct \approx R \tag{16}$$

which is again the Machian relation. It says that the gravitational radius of the Universe is of the order of its size, or that the relativistic energy of any mass m is of the order of its gravitational potential energy, with respect to the rest of the Universe.

## 4. The Einstein's Cosmological Equations

The Einstein cosmological equations are, with the equivalent notation $a \equiv R$:

$$\left(\frac{\dot{a}}{a}\right)^2 + \frac{2\ddot{a}}{a} + 8\pi G \frac{p}{c^2} + \frac{kc^2}{a^2} = \Lambda c^2 \tag{17}$$

$$\left(\frac{\dot{a}}{a}\right)^2 - \frac{8\pi}{3} G\rho + \frac{kc^2}{a^2} = \frac{\Lambda c^2}{3} \tag{18}$$

Since H is equal to $\dot{a}(t)/a(t)$ and $p = w\rho_p c^2$, the barotropic energy density, we can divide both equations by $H^2$ to get

$$1 + 2\frac{a''a}{a'^2} + 3w\Omega_p + \Omega_k = 3\Omega_\Lambda \tag{19}$$

$$1 - \Omega_m + \Omega_k = \Omega_\Lambda \tag{20}$$

where $\Omega_p$ is the parameter related to pressure that we define as:

$$\Omega_p = \frac{8\pi}{3}\frac{G\rho_p}{H^2} = \Omega_m \frac{\rho_p}{\rho} \tag{21}$$

Now, locally $\Omega_k \approx 0$, so that we get the solutions for (19) and (20) found experimentally: $\Omega_m \approx 1/3$, $\Omega_\Lambda \approx 2/3$. With $a'' \approx 0$ we get the condition:

$$3w\Omega_p \approx 1 \tag{22}$$

With the local measurement of w ≈ -1 one gets the result:

$$\Omega_p \approx -\frac{1}{3} \tag{23}$$

This means that approximately it is

$$\Omega_m + \Omega_p \approx 0 \tag{24}$$

The local interpretation when observing the Universe is that it is expanding almost linearly with time. This is consistent with a local interpretation of a constant speed of light $c_0$ that gives an expansion rate as $a(t) \approx c_0 t$. However, we note that the cosmological point of view, as opposed to the local one, defers because cosmologically it is $a(t) = ct = constant$. In this case the solution to the Einstein's cosmological equations from this point of view must have $a''(t) = a'(t) = 0$, and k = 1.

**5. Integration of the Bianchi Identity**

We have derived elsewhere, Belinchón and Alfonso-Faus (2001), the expression for the zero value of the right hand side of the Einstein's field equations:

$$\nabla(\frac{8\pi G}{c^4}T^{\mu\nu} + \Delta g^{\mu\nu}) = 0 \tag{25}$$

which is now

$$\frac{\rho'}{\rho} + 3(\omega\frac{\rho_p}{\rho} + 1)H + \frac{\Lambda' c^4}{8\pi G\rho} + \frac{G'}{G} - 4\frac{c'}{c} = 0 \qquad (26)$$

Here we have allowed for G, c and Λ to be time dependent, if such is the case. The cosmological parameter Λ is a real constant so that it disappears from (26). Here ρ is the energy density and integration of (26) gives,

$$\frac{G\rho}{c^4} R^{3(1-\omega\frac{\rho_p}{\rho})} = const \qquad (27)$$

or equivalently

$$\frac{GM}{c^2} R^{-3\omega\frac{\rho_p}{\rho}} = const \qquad (28)$$

This result is very important. We see that there is agreement between this theoretical result and the interpretation of Mach's principle given by $GM/c^2 \approx R$, as in equation (16), provided that the product $3w\rho_p/\rho$ be exactly one:

$$3w\rho_p / \rho = 1 \qquad (29)$$

We get that R is constant from the condition that all the gravitational radii of masses M inside their proper volume are constant, Alfonso-Faus (2008). The cosmological point of view has R = constant, because R = ct is constant, and all the gravitational radii are constants. For this case the equations (17) and (18) and (29) reduce to:

$$\frac{8\pi}{3} G\rho + \frac{kc^2}{a^2} = \Lambda c^2 \qquad (30)$$

$$-\frac{8\pi}{3} G\rho + \frac{kc^2}{a^2} = \frac{\Lambda c^2}{3} \qquad (31)$$

In terms of the Ω parameters these two equations become:

$$\Omega_\rho + \Omega_k = 3\Omega_\Lambda$$
$$-\Omega_\rho + \Omega_k = \Omega_\Lambda \qquad (32)$$

Since R ≈ ct one has $\Omega_k$ ≈ 1 and from (32) we get $\Omega_\rho$ ≈ $\Omega_\Lambda$ ≈ ½, an approximate equipartition of energy.

**6. The Cosmological Constant Problem. Cosmic Planck's Constant**

The cosmological constant Λ, that we have determined to be a true constant in (8), has an approximate numerical value of $2 \times 10^{-56}$ cm$^{-2}$. The corresponding energy density is of the following order of magnitude:

$$\rho_\Lambda \approx \frac{\Lambda c^4}{8\pi G} = \frac{c^4}{4\pi L^2 G} = const \; x \; c = const/t \qquad (33)$$

where we have used (8) and the ratio $G/c^3$ = constant. The equivalent mass density has the value to day:

$$\rho_\Lambda / c^2 \approx \frac{c^2}{4\pi L^2 G} = 2.4 \times 10^{-29} \; gr/cm^3 \qquad (34)$$

It is proportional to time. In many cases this is interpreted as the equivalent mass density of vacuum. If we look at the Planck's mass density, where we can consider the fluctuations of vacuum due to the Planck's quantum black holes, we get:

$$\rho_p / c^2 = \frac{\left(\frac{\hbar c}{G}\right)^{1/2}}{\frac{4\pi}{3}\left(\frac{G\hbar}{c^3}\right)^{3/2}} = \frac{3}{4\pi}\frac{c^5}{\hbar G^2} \approx 10^{93} \; gr/cm^3 \qquad (35)$$

which is also proportional to time. Then the ratio between (35) and (34) is a universal constant:

$$\frac{\rho_p}{\rho_\Lambda} = 3\frac{c^3 L^2}{G\hbar} = 3\left(\frac{L}{l_p}\right)^2 \approx 10^{122} \tag{36}$$

This is a large number indeed. It pops up in many instances in cosmology. It is not very convincing that such a big difference between the two mass densities is real at all. So, there is a strong feeling that we are doing something wrong. The best guess is that the two mass densities in (36) are of the same order of magnitude. But, which is the right way to lower such a big number? We maintain the validity of the Einstein field equations, as derived from the action principle. Then $G/c^3$ is a true universal constant that can be taken as unity in a certain system of units. Also M = t, the Mass-Boom, Alfonso-Faus (2008). On the other hand the constant length L can also be taken as unity, in the same system. It is the product of c times t, a universal constant. These two conditions define the unit of length and identify mass and time as one and the same thing. Since mass is quantized so is time. We are left to conclude that Planck's constant ℏ in (36) has to be of the order of $3 \times 10^{-121}$ in these units.

Planck's constant is of the order of the product of mc, m the mass of a fundamental particle, times the size $r_p$ of such a particle. If this particle contributes significantly to the mass of the Universe, then their number $N_p$ is of the order of $10^{80}$ and the ratio of the size of the Universe to the size of the particle is about $10^{41}$. The product of these two numbers is the right one to convert Planck's constant to a cosmological Planck's constant $\hbar_c$ of order one:

$$\hbar_c \approx N_p mc \cdot (ct) \approx 1 \tag{37}$$

This should be the right Planck's constant to be used in cases of cosmological quantum physics. It converts the large number in (36) into a number of order one. The same result is obtained when using Planck's units. Planck's constant is now given by $\hbar \approx m_* \, c \, l_* \approx 10^{-122}$. One converts this relation for a cosmological scale by multiplying $m_*$ and $l_*$ by $10^{61}$. Then we get again $\hbar_c \approx Mc.L \approx 1$.

**8. The entropy problem**

There are many different ways to express the entropy of a system. Dealing with the Universe, and considering it as a black hole, we can apply the Hawking formulation n (1975):

$$S_g \approx \frac{k}{\hbar c} GM^2 = \frac{k}{\hbar} \approx 10^{122} k \qquad (38)$$

where we have used $G = c^3$ and $Mc = 1$. The Bekenstein (1972) upper limit is given by:

$$S_g \approx k \frac{ER}{\hbar c} = \frac{k}{\hbar} \approx 10^{122} k \qquad (39)$$

where E is the energy of the Universe and $R = ct = L = 1$, its size. We can define this gravitational entropy as the number of gravity quanta in the Universe, Alfonso-Faus (1999):

$$S_g \approx k \frac{M}{m_g} = k \frac{c^2 t}{\hbar} M = \frac{k}{\hbar} \approx 10^{22} k \qquad (40)$$

And finally we can make use of the thermodynamic definition, with the temperature $T_g$ of the gravity quanta:

$$S_g \approx \frac{Mc^2}{T_g} = \frac{c}{T_g} \approx \frac{k}{\hbar} \approx 10^{122} k \qquad (41)$$

$$T_g \approx \frac{\hbar c}{k} = \frac{\hbar}{k} \frac{1}{t} \qquad (42)$$

that varies inversely proportional to the age of the Universe t. If we take the entropy as defined by the number of parts a system consist of, then there must be $10^{122}$ gravity quanta in the Universe. On the other hand, if we use the cosmological Planck's constant $\hbar_c \approx 1$ instead of the Planck's constant, all the definitions of the gravitational entropy give $S = k$, and $T_g \approx c$ (with $k = 1$). The gravitational entropy of the Universe turns out to be just the Boltzmann constant, which seems a plausible result for the whole Universe. The horizon problem is solved because ct = constant and the Universe is causally connected always.

For the photons in the Universe the entropy is $N_{ph} \approx 10^{87}$ k which has a thermodynamic definition as

$$S_{ph} \approx \frac{N_{ph}\hbar c/\lambda}{T_{ph}} \approx 10^{87}\frac{\hbar c}{T_{ph}}10^{29} = 10^{17}\frac{\hbar}{tT_{ph}} \approx 10^{87}k \qquad (43)$$

$$T_{ph} \approx 10^{29}\frac{\hbar}{kt} \qquad (44)$$

The photon temperature is then $10^{29}$ times the gravity quanta temperature (42). This number is the inverse of the photon wavelength, with ct = L = 1. Then the gravity quanta wavelength is of the order of L, as it should, Alfonso-Faus[10]. The consequence of Einstein's equivalence principle is that gravity cannot be detected at any point. The energy-momentum of gravity cannot have a proper local density, it is fundamentally non-local, Misner, Thorne and Wheeler (1973). This property is ensured by forcing the wavelength of the gravity quanta to be of the order of the size of the Universe L. Its mass $m_g$ is then $\hbar/Lc = \hbar/c^2 t$ as we have used in (40).

## 9. Cosmological Quantum Physics: Schrödinger equation

A standard technique for the solution of the Schrödinger equation is to look for solutions which are products of a function of x, $\psi(x)$, times a function of time t, $\phi(t)$. Since we have found that energies vary inversely proportional to t, the Schrödinger equation multiplied by t becomes:

$$\frac{t}{\psi(x)}\left[-\frac{\hbar^2}{2m}\frac{d^2\psi(x)}{dx^2}+V(x)\psi(x)\right] = i\hbar\frac{t}{\phi(t)}\frac{d\phi(t)}{dt} = Et = const. \qquad (45)$$

The analysis of this equation gives a straightforward result: the spatial part does not have any change from the well known solutions. This is because the ratio t/m is a constant (mass-boom), ant the product tV(x) does not have the time t in it (the potential energy varies as 1/t). Hence, the time independent equation is the usual one: an ordinary second order differential equation, closely related to the time independent differential equation for classical wave motion. The functions $\psi(x)$ are the *eigenfunctions*. The time dependent equation is different: integration of the second part of (41) gives the solution

$$\phi(t) = const. \; t^{\frac{Et}{i\hbar}} \tag{46}$$

It is interesting to see that if we use an imaginary time, t = it' then the exponent of t in (46) is real, Et'/ℏ. If one uses universal values, E = Mc² = c, ℏ$_c$ ≈ 1 and ct' = 1, we get again the same solution that we obtained from the Einstein's cosmological equations: a linear expansion of the Universe. General relativity and quantum mechanics give the same local solution of an expanding Universe with the law *a(t) = const. t*. Obviously this constant is just the speed of light today.

**10. Conclusions**

The idea of Dirac (1937) that there should not be any large number in cosmology is implemented here. We define a cosmological Planck's constant ℏ$_c$ that is about $10^{120}$ times larger than the usual ℏ. Then all cosmological dimensionless parameters are of order one.

Since the speed of light must be proportional to the Hubble parameter H, it decreases linearly with cosmological time, the same as H. The next conclusion is that Zel'dovich's expression and Weinberg's relation is one and the same thing. Since we have noted that this last relation is a result of the combined application of the conservation of momentum and the universal gravitation (Newton's laws), so is the Zel'dovich's expression. It implies that the cosmological constant Λ is a true universal constant. All the cosmological Ω's parameters are constant.

Locally Ω$_k$ ≈ 0, and we get the solutions found experimentally, Ω$_Λ$ = 2/3, Ω$_m$ = 1/3, and the value w = -1. The barotropic parameter has the value Ω$_p$ = -1/3. The local interpretation when observing the Universe is that it is expanding almost linearly with time. This is consistent with a local interpretation of a constant speed of light c$_0$ that gives an expansion rate as *a(t) = c$_0$t*.

The cosmological point of view is: *a* = R = constant because ct is constant. And all the gravitational radii are constants.

The cosmological constant problem is avoided here by the introduction of the cosmic Planck's constant $\hbar_c \approx N_p mc \cdot (ct) \approx 1$. This should be the right Planck's constant to be used when dealing with cosmological quantum physics.

If we want to obtain the present mass, length and age of the Universe out of a Planck's fluctuation we need a spatial "inflation" and a mass-boom (and time-boom) by the same factor of order $10^{61}$. This converts Planck's fluctuation into the cosmological Planck's fluctuation (the Universe today): general relativity and quantum mechanics give the same local expansion law for the Universe, a linear one. It is suggested that the constant number $10^{61}$ is representative of the past inflation phase.

If we use the cosmological Planck's constant $\hbar_c \approx 1$ instead of the Planck's constant, all the definitions of the gravitational entropy give $S \approx k$, and $T_g \approx c/k = 1/kt$, the temperature of gravity quanta. The gravitational entropy of the Universe turns out to be just k, the Boltzmann constant, which seems a plausible result for the whole Universe. The horizon problem disappears because ct = constant and the Universe is always causally connected.

We advance one more step in cosmological quantum physics. General relativity and quantum mechanics give here the same local solution of an expanding Universe with the law *a(t) = const. t*. This constant is just the speed of light today.

## 11. References

Alfonso-Faus, A., (2008), "The Speed of Light and the Hubble Parameter: The Mass-Boom Effect", arXiv:0710.3039. Accepted to appear in *Astrophysics and Space Science* Journal, March 2008.

Alfonso-Faus, A., (2000), "Gravity quanta, entropy and black holes". Phys. Essays 12, 673, (1999) and arXiv.org gr-qc/0002065 (2000).

Bekenstein, J., (1972), J. Phys. Rev., D7, 2333-2346

Belinchón, J.A. and Alfonso-Faus, A., (2001), Int. J. Mod. Phys., "A Theory of Time-Varying Constants", D10, 299-310 and arXiv:gr-qc/0404044.

Dirac, P.A.M., (1937) Nature, **139,** 323, 1937, and Proc. Roy. Soc. London, **A 165**: 199-208, 1938.

Eisberg, R.M., (1961), *"Fundamentals of Modern Physics",* John Wiley & Sons, Inc.

Hawking, S.W., (1975), Commun. Math. Phys., 43, 199-220.

Misner, C. W., Thorne, K. S. & Wheeler, J. A., (1973). *"Gravitation"* (Freeman, San Francisco).

Uzan, J.P., (2003), Reviews of Modern Physics, **75**:403-455, and hep-ph/0205340.

Weinberg, S., (1972), *Gravitation and Cosmology*, John Wiley and Sons, New York.

Zel'dovich, Ya.B., Pis'ma Zh. Eksp. Teor. Fiz., (1967), vol, **6**, p. 883.